\documentclass[aps,prl,twocolumn,showpacs,preprintnumbers,amsmath,amssymb]{revtex4}
\usepackage{graphicx}
\usepackage{dcolumn}
\usepackage{bm}

\begin{document}
\title{High mobility two-dimensional electron system on \\
hydrogen-passivated silicon(111) surfaces}

\author{K. Eng}
\email{kevin@lps.umd.edu}
\author{R. N. McFarland}%
\author{B. E. Kane}
\affiliation{Laboratory for Physical Sciences, University of Maryland at College Park,
College Park, MD 20740}%

\date{\today}

\begin{abstract}

We have fabricated and characterized a field-effect transistor in
which an electric field is applied through an encapsulated vacuum
cavity and induces a two-dimensional electron system on a
hydrogen-passivated Si(111) surface.  This vacuum cavity preserves
the ambient sensitive surface and is created via room temperature
contact bonding of two Si substrates.  Hall measurements are made
on the H-Si(111) surface prepared in aqueous ammonium fluoride
solution.  We obtain electron densities up to $6.5 \times 10^{11}$
cm$^{-2}$ and peak mobilities of $\sim 8000$ cm$^{2}$/V s at 4.2
K.

\end{abstract}

\maketitle

    Electron inversion or accumulation layers on a low disorder semiconductor
surface can potentially be a new high quality two-dimensional
electron system (2-DES) and could present a new technique in the
development of atomic-scale electronic devices, where electrons
are coupled to single atoms and molecules positioned on the
surface.  However, creating a 2-DES on a semiconductor surface is
non-trivial due to the usual presence of dangling bonds and
surface contamination. Such disorder creates surface states
(within the semiconductor's band gap) which may trap charge
carriers.  An ideal surface would be clean of contamination and
have all its dangling bonds passivated.

    Initial studies\cite{weinberger,yabl} of Si(111) surfaces chemically treated
with hydrofluoric acid (HF) established nearly ideal electronic
properties: these surfaces exhibit both low surface-recombination
velocity and low surface state densities ($\leq 10^{10}$
cm$^{-2}$)\cite{weinberger}. Subsequent infrared absorption
measurements\cite{yabl,higashi} showed that the dangling bonds on
the Si surface are passivated by hydrogen. It was also discovered
that buffering HF with ammonium fluoride (NH$_4$F) produces a more
anisotropic Si etchant, where the (111) planes etch more slowly
than other crystallographic planes. The Si(111) surface is thus
unique in that such a simple wet chemical treatment can produce an
ideal H-passivated Si surface which is atomically flat and defect
free\cite{higashi}. In addition, the H-Si(100) surface has been
utilized recently as a resist for controlling the placement of
individual atoms\cite{donors,qubit}. Unfortunately H-passivated
surfaces are sensitive to ambient conditions (oxidizes $\sim$
hours) and elevated temperatures (T $\geq 300^{\circ}$C). Hence,
previous measurements\cite{weinberger,yabl,gerko} on H-Si surfaces
have been constrained to controlled environments, and to date this
surface has not been incorporated into practical devices. However,
preservation of the H-passivated surface has been demonstrated by
contact bonding two H-Si(111) substrates together at room
temperature\cite{grey}. In this letter we describe the fabrication
of a H-passivated Si field effect transistor (FET) which allows a
2-DES to be gated on a H-Si(111) surface through an encapsulated
vacuum cavity, and we report the first electron transport
measurements on H-Si(111) surfaces at 4.2 K.

\begin{figure}
\begin{center}
 \includegraphics[height=2.9in,width=3in,clip]{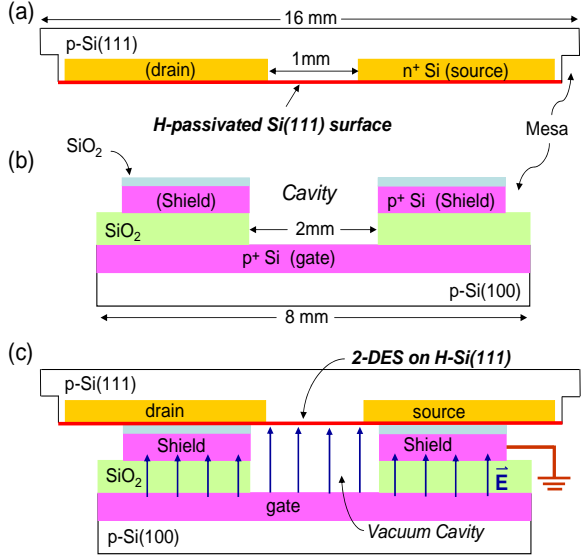}
  \caption{Schematic cross-section of the Si(111) and SOI
  substrates before and after bonding.
  (a) The H-Si(111) substrate has a mesa etched around the edges and
  n$^+$ contact regions made through P implantation.
  (b) Within the SOI substrate, the gate and shield layers are created by a double B
  implantation which are represented by the two p$^+$ doped Si layers.  RIE is then used to
  etch a mesa and a cavity.  (c) A H-Si(111) FET, where both substrates
  ((a) \& (b)) are contact-bonded in vacuum.
  The blue arrows depict the electric field produced by the gate
  which is terminated at the shield layer except inside the vacuum
  cavity, where it induces a 2-DES on the H-Si(111) surface.
  } \label{substrates}
\end{center}
\end{figure}

    Fabrication of the device begins with two individual Si
substrates ($\sim$ 1 cm$^2$), each having a distinct function. One
is the H-passivated Si(111) substrate with source and drain
contacts (Fig.~\ref{substrates}a). It is at this surface where the
2-DES will be created.  The other is a silicon-on-insulator (SOI)
substrate which acts as the ``remote gate''
(Fig.~\ref{substrates}b), where an electric field can be
controlled within an etched cavity.  These two substrates are
contact-bonded in vacuum, and they adhere to one another due to
local van der Waals forces between the two mating surfaces.
Successful bonding requires these surfaces to be \textit{flat} and
\textit{clean} of particle contamination before
contact\cite{waferbondingbook}.  The bonding not only creates the
FET, where the ``remote gate'' can induce electrons on the
H-Si(111) surface, but also encapsulates the air sensitive surface
in a vacuum cavity (Fig.~\ref{substrates}c).

\begin{figure}
\begin{center}
  \includegraphics[height=2.55in,width=3.0in,clip]{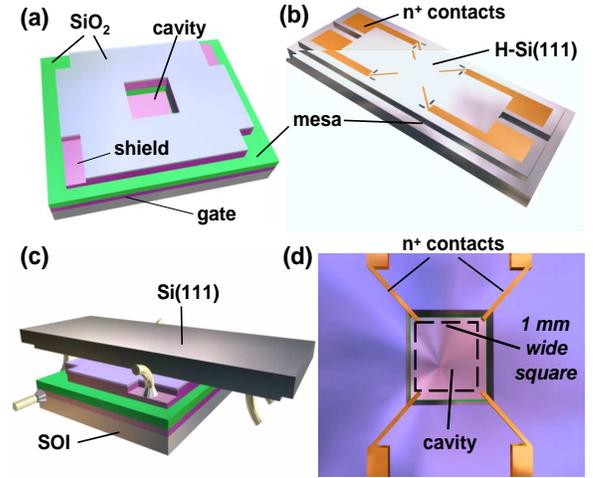}
  \caption{Diagram showing the assembly of an encapsulated vacuum H-Si(111)
  FET. (a) The SOI substrate has a mesa, shield contacts, and a cavity etched via RIE.
  (b) The H-Si(111) substrate with its mesa and four n$^+$ contacts.
  (c) Shows the H-Si(111) substrate (b) flipped over and
  bonded to the SOI.  The conducting regions are then soldered with In and Au wires.
  (d) Top view of (c) the bonded H-Si(111) FET, where the H-Si(111) substrate
  is drawn to be transparent in order to show the alignment of the n$^+$
  contacts with respect to the cavity.  }
\label{crosssection}
\end{center}
\end{figure}

    The H-passivated Si surface is fabricated on a 16 mm $\times$ 7 mm
p-type Si(111) substrate (FZ, $\rho \sim$ 200 $\Omega$ cm, $\leq$
0.2$^\circ$ miscut).   First, a 5 $\mu$m deep Si mesa is
dry-etched around the edges of the sample (Fig.~\ref{substrates}a
\& Fig.~\ref{crosssection}b) along with alignment marks via
reactive ion etching (RIE). This mesa prevents particles from
accumulating on the surface while handling the substrate and
ensures a clean edge for bonding.  Next, a sacrificial SiO$_2$
layer ($\sim$ 300 \AA) is thermally grown on top of the Si(111)
substrate for ion implantation. Photoresist is applied and
patterned to expose four contact regions where 50 keV P ions (dose
of $4.5 \times 10^{14}$ cm$^{-2}$) are implanted through the
SiO$_2$. The implantation is activated by a 30 min 950$^{\circ}$C
anneal in dry N$_2$.  Since the oxidation and etch rates of Si
depend strongly on the doping level, these implantation and
annealing parameters are optimized in such a way as to minimize
surface topography differences around the edges of the implanted
regions and at the same time maintain a contact resistance ($\rho
\sim 100$ $\Omega$/$\Box$ at T = 4.2 K) which is still metallic.
As shown in Fig.~\ref{crosssection}d, each of the four n$^+$
contacts approaches a corner of a 1 mm wide square at the center
of the Si(111) substrate in order to facilitate Van der Pauw
measurements.

    Fabrication of the ``remote gate'' starts with a 8 mm $\times$ 12 mm
p-type SOI substrate (SOITEC UNIBOND), which consists of 3400 \AA \
thick p-Si(100) (FZ, $\rho >$ 2000 $\Omega$ cm) on top of a 4000 \AA
\ thick insulating SiO$_2$ film.  First, two conducting layers are
created by B implantation in both silicon layers within the SOI: 250
keV at $4 \times 10^{14}$ cm$^{-2}$ and 100 keV at $2 \times
10^{14}$ cm$^{-2}$. As shown in Fig.~\ref{substrates}b, these
degenerately p$^+$ doped Si layers form the gate and shield
respectively.  The purpose of the shield is to restrict the action
(electric field) of the gate to a well defined region exposed by an
etched cavity within the SOI (Fig.~\ref{substrates}c).  The B
dopants are annealed at 950$^{\circ}$C for 30 min in dry O$_2$,
which adds a 250 \AA \ SiO$_2$ layer on top of the SOI. This SiO$_2$
layer electrically isolates the shield from the n$^+$ contacts on
the H-Si(111) substrate. The SOI substrate then goes through a
series of three lithography steps using RIE. A mesa is defined
around the edges of the substrate by etching away the top SiO$_2$
and Si layer of the SOI.  Contacts to the shield layer are then
exposed by etching \ $\sim$ 2700 \AA \ of p-Si. Lastly, a $2 \times
2$ mm$^2$ cavity (depth $\sim$ 7600 \AA) at the center of the SOI
substrate is etched to the gate layer (Fig.~\ref{substrates}b \&
Fig.~\ref{crosssection}a).

    Before bonding, both Si(111) and SOI substrates are cleaned
of particles and organic contamination.
The sacrificial SiO$_2$ film on top of the Si(111) substrate is
removed in 10\% HF and then the surface is H-passivated by
immersion in a N$_2$ sparged (40\%) NH$_4$F aqueous
solution\cite{wade} for 4 min.

\begin{figure}
\begin{center}
  \includegraphics[height=2.35in,width=3.1in,clip]{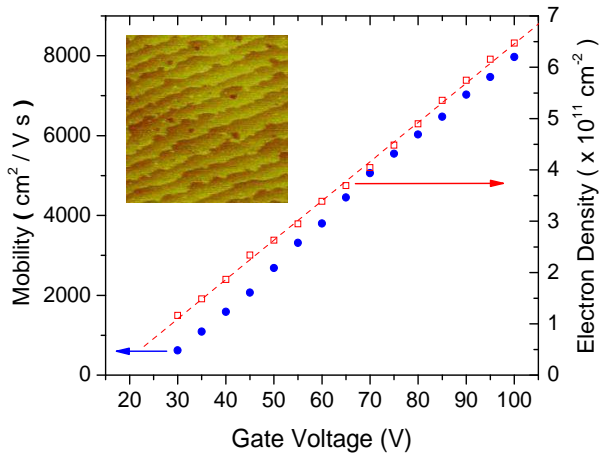}
  \caption{Hall measurements of the electron density and
  mobility vs. gate voltage at 4.2 K.  The red dashed line is a linear fit of the data.
  Inset: AFM image
of a $2 \times 2$ $\mu$m$^2$ H-Si(111) surface after 4 min in
N$_2$ sparged 40\% NH$_4$F solution.}
  \label{density}
\end{center}
\end{figure}

   Since the flatness of the H-Si(111) and SOI surfaces is
critical for bonding and encapsulating the H-Si(111), both mating
surfaces were investigated through (\textit{ex situ}) atomic force
microscopy (AFM). The inset of Fig.~\ref{density} shows an AFM
image of a H-Si(111) surface prepared in N$_2$-sparged NH$_4$F
solution, where the atomic step widths agree with the wafer miscut
angle.  The average rms roughness of the H-Si(111) surface is 1.5
\AA \ and the SiO$_2$ surface (on the SOI) is $\leq$ 2 \AA.
Although the micro-roughness of these surfaces makes them suitable
for bonding\cite{waferbondingbook}, fabricating an atomically flat
H-Si(111) surface with electrical contacts
(Fig.~\ref{substrates}a) is non-trivial due to inherent surface
height differences between the n$^+$ contact regions and the
surrounding p-type Si(111).  Our implantation and NH$_4$F etching
parameters have minimized such height differences to $\leq$ 5\AA.

    Within $\sim5$ min of preparing the
H-Si(111) surface, both substrates are
placed in vacuum of a base pressure of $\leq 10^{-6}$ Torr. They
are then contacted at room temperature, and visual confirmation of
the bonding is made through an infrared camera (similar to
Fig.~\ref{crosssection}d). The bond is then tempered in vacuum at
100$^{\circ}$C for $\sim$ 12 hours.

    After tempering the bond, the encapsulated H-Si(111) FET is
characterized by Hall mobility and density measurements at 4.2 K
using standard Van der Pauw techniques. Contacts to the shield
layer are grounded throughout the measurements.  Hall data from a
representative H-Si(111) FET, shown in Fig.~\ref{density}, follows
theoretical predictions\cite{ando2} of a parallel-plate capacitor
where the electron density on the H-Si(111) surface is linearly
dependent on the gate voltage.  From the linear fit shown in
Fig.~\ref{density}, the dielectric of the cavity is calculated to
be within 5\% of the expected value for a vacuum cavity with our
dimensions.  Due to the device architecture, the maximum electron
density which can be induced on the H-Si(111) surface is limited
by the dielectric breakdown voltage of the 4000 \AA \ SiO$_2$
layer within the SOI. This breakdown voltage can be as high as 130
V, but the device in Fig.~\ref{density} has a peak density of
$\sim 6.5 \times 10^{11}$ cm$^{-2}$ at 100 V.

    Because inversion layers in Si are thin ($\sim$ 100 \AA),
the electron mobility is sensitive to scattering associated with
the surface\cite{ando2} and thus provides a simple tool for
analyzing the quality of our prepared H-Si(111) surfaces.  The
peak electron mobility on our H-Si(111) surface is $\sim8000$
cm$^{2}$/V s at 4.2 K (Fig.~\ref{density}). Compared to previous
peak electron mobility measurements in Si(111) metal oxide
semiconductor (MOS) FETs ($\sim 2500$ cm$^{2}$/V
s)\cite{tsui,cole}, this is the highest electron mobility recorded
on a Si(111) surface.  In conjunction with the high mobility, the
threshold of this device is 15 V, which indicates that the number
of trapped electrons is $\sim 10^{11}$ cm$^{-2}$ at 4.2 K.  If we
assume these are due to surface states arising from dangling bonds
on the Si(111) surface, then it demonstrates that our chemically
prepared surfaces (in a cleanroom environment) remain H-passivated
during measurements.  It is also interesting to note that the
mobility is linear with density above $3.5 \times 10^{11}$
cm$^{-2}$.



    In summary, we have presented a new technique for gating
air sensitive surfaces or materials through an encapsulated vacuum
cavity.  Utilizing vacuum as a gate dielectric, we have fabricated
a 2-DES on a H-Si(111) surface which exhibits higher mobilities
than previous Si(111) MOSFETs.  Further electron transport
measurements on such H-Si FETs can potentially make important
contributions to 2-D physics.  In addition, this experimental
technique has potential toward the development of atomic-scale
electronic devices, where electrons are coupled to specific
molecules or single atoms positioned on the hydrogen-passivated Si
surface

    This research was funded by the National Security Agency and the Advanced
Research and Development Activity.

\bibliographystyle{apsrev}

\begin{thebibliography}{20}
\expandafter\ifx\csname
natexlab\endcsname\relax\def\natexlab#1{#1}\fi
\expandafter\ifx\csname bibnamefont\endcsname\relax
  \def\bibnamefont#1{#1}\fi
\expandafter\ifx\csname bibfnamefont\endcsname\relax
  \def\bibfnamefont#1{#1}\fi
\expandafter\ifx\csname citenamefont\endcsname\relax
  \def\citenamefont#1{#1}\fi
\expandafter\ifx\csname url\endcsname\relax
  \def\url#1{\texttt{#1}}\fi
\expandafter\ifx\csname
urlprefix\endcsname\relax\def\urlprefix{URL }\fi
\providecommand{\bibinfo}[2]{#2}
\providecommand{\eprint}[2][]{\url{#2}}

\bibitem[{\citenamefont{Weinberger et~al.}(1985)\citenamefont{Weinberger,
  Deckman, Yablonovitch, Gmitter, Kobasz, and Garoff}}]{weinberger}
\bibinfo{author}{\bibfnamefont{B.~R.} \bibnamefont{Weinberger}},
  \bibinfo{author}{\bibfnamefont{H.~W.} \bibnamefont{Deckman}},
  \bibinfo{author}{\bibfnamefont{E.}~\bibnamefont{Yablonovitch}},
  \bibinfo{author}{\bibfnamefont{T.}~\bibnamefont{Gmitter}},
  \bibinfo{author}{\bibfnamefont{W.}~\bibnamefont{Kobasz}}, \bibnamefont{and}
  \bibinfo{author}{\bibfnamefont{S.}~\bibnamefont{Garoff}},
  \bibinfo{journal}{J. Vac. Sci. Technol.} \textbf{\bibinfo{volume}{3}},
  \bibinfo{pages}{887} (\bibinfo{year}{1985}).

\bibitem[{\citenamefont{Yablonovitch et~al.}(1986)\citenamefont{Yablonovitch,
  Allara, Chang, Gmitter, and Bright}}]{yabl}
\bibinfo{author}{\bibfnamefont{E.}~\bibnamefont{Yablonovitch}},
  \bibinfo{author}{\bibfnamefont{D.~L.} \bibnamefont{Allara}},
  \bibinfo{author}{\bibfnamefont{C.~C.} \bibnamefont{Chang}},
  \bibinfo{author}{\bibfnamefont{T.}~\bibnamefont{Gmitter}}, \bibnamefont{and}
  \bibinfo{author}{\bibfnamefont{T.~B.} \bibnamefont{Bright}},
  \bibinfo{journal}{Phys. Rev. Lett.} \textbf{\bibinfo{volume}{57}},
  \bibinfo{pages}{249} (\bibinfo{year}{1986}).

\bibitem[{\citenamefont{Higashi et~al.}(1990)\citenamefont{Higashi, Chabal,
  Trucks, and Raghavachari}}]{higashi}
\bibinfo{author}{\bibfnamefont{G.~S.} \bibnamefont{Higashi}},
  \bibinfo{author}{\bibfnamefont{Y.~J.} \bibnamefont{Chabal}},
  \bibinfo{author}{\bibfnamefont{G.~W.} \bibnamefont{Trucks}},
  \bibnamefont{and}
  \bibinfo{author}{\bibfnamefont{K.}~\bibnamefont{Raghavachari}},
  \bibinfo{journal}{Appl. Phys. Lett.} \textbf{\bibinfo{volume}{56}},
  \bibinfo{pages}{656} (\bibinfo{year}{1990}).

\bibitem[{\citenamefont{Shen et~al.}(1997)\citenamefont{Shen, Wang, and
  Tucker}}]{donors}
\bibinfo{author}{\bibfnamefont{T.-C.} \bibnamefont{Shen}},
  \bibinfo{author}{\bibfnamefont{J.}~\bibnamefont{Wang}}, \bibnamefont{and}
  \bibinfo{author}{\bibfnamefont{J.~R.} \bibnamefont{Tucker}},
  \bibinfo{journal}{Phys. Rev. Lett.} \textbf{\bibinfo{volume}{78}},
  \bibinfo{pages}{1271} (\bibinfo{year}{1997}).

\bibitem[{\citenamefont{O'Brien et~al.}(2001)\citenamefont{O'Brien, Schofield,
  Simmons, Clark, Dzurak, Curson, Kane, McAlpine, Hawley, and Brown}}]{qubit}
\bibinfo{author}{\bibfnamefont{J.~L.} \bibnamefont{O'Brien}},
  \bibinfo{author}{\bibfnamefont{S.~R.} \bibnamefont{Schofield}},
  \bibinfo{author}{\bibfnamefont{M.~Y.} \bibnamefont{Simmons}},
  \bibinfo{author}{\bibfnamefont{R.~G.} \bibnamefont{Clark}},
  \bibinfo{author}{\bibfnamefont{A.~S.} \bibnamefont{Dzurak}},
  \bibinfo{author}{\bibfnamefont{N.~J.} \bibnamefont{Curson}},
  \bibinfo{author}{\bibfnamefont{B.~E.} \bibnamefont{Kane}},
  \bibinfo{author}{\bibfnamefont{N.~S.} \bibnamefont{McAlpine}},
  \bibinfo{author}{\bibfnamefont{M.~E.} \bibnamefont{Hawley}},
  \bibnamefont{and} \bibinfo{author}{\bibfnamefont{G.~W.} \bibnamefont{Brown}},
  \bibinfo{journal}{Phys. Rev. B} \textbf{\bibinfo{volume}{64}},
  \bibinfo{pages}{161401} (\bibinfo{year}{2001}).

\bibitem[{\citenamefont{Oskam et~al.}(1996)\citenamefont{Oskam, Hoffmann, and
  Searson}}]{gerko}
\bibinfo{author}{\bibfnamefont{G.}~\bibnamefont{Oskam}},
  \bibinfo{author}{\bibfnamefont{P.~M.} \bibnamefont{Hoffmann}},
  \bibnamefont{and} \bibinfo{author}{\bibfnamefont{P.~C.}
  \bibnamefont{Searson}}, \bibinfo{journal}{Phys. Rev. Lett.}
  \textbf{\bibinfo{volume}{76}}, \bibinfo{pages}{1521} (\bibinfo{year}{1996}).

\bibitem[{\citenamefont{Grey and Hermansson}(1997)}]{grey}
\bibinfo{author}{\bibfnamefont{F.}~\bibnamefont{Grey}} \bibnamefont{and}
  \bibinfo{author}{\bibfnamefont{K.}~\bibnamefont{Hermansson}},
  \bibinfo{journal}{Appl. Phys. Lett.} \textbf{\bibinfo{volume}{71}},
  \bibinfo{pages}{3400} (\bibinfo{year}{1997}).

\bibitem[{\citenamefont{Gosele and Tong}(1999)}]{waferbondingbook}
\bibinfo{author}{\bibfnamefont{U.}~\bibnamefont{Gosele}} \bibnamefont{and}
  \bibinfo{author}{\bibfnamefont{Q.~Y.} \bibnamefont{Tong}},
  \emph{\bibinfo{title}{Semiconductor Wafer Bonding: Science and Technology}}
  (\bibinfo{publisher}{John Wiley \& Sons Inc}, \bibinfo{year}{1999}).


\bibitem[{\citenamefont{Wade and Chidsey}(1997)}]{wade}
\bibinfo{author}{\bibfnamefont{C.~P.} \bibnamefont{Wade}} \bibnamefont{and}
  \bibinfo{author}{\bibfnamefont{C.~E.~D.} \bibnamefont{Chidsey}},
  \bibinfo{journal}{Appl. Phys. Lett.} \textbf{\bibinfo{volume}{71}},
  \bibinfo{pages}{1679} (\bibinfo{year}{1997}).

\bibitem[{and()}]{ando2}
\bibinfo{note}{See T. Ando, A. B. Fowler, and F. Stern, Rev. Mod. Phys.
  \textbf{54}, 437, (1982). and references therein}.

\bibitem[{\citenamefont{Tsui and Kaminsky}(1976)}]{tsui}
\bibinfo{author}{\bibfnamefont{D.~C.} \bibnamefont{Tsui}} \bibnamefont{and}
  \bibinfo{author}{\bibfnamefont{G.}~\bibnamefont{Kaminsky}},
  \bibinfo{journal}{Solid State Comm.} \textbf{\bibinfo{volume}{20}},
  \bibinfo{pages}{93} (\bibinfo{year}{1976}).

\bibitem[{\citenamefont{Cole and McCombe}(1984)}]{cole}
\bibinfo{author}{\bibfnamefont{T.}~\bibnamefont{Cole}} \bibnamefont{and}
  \bibinfo{author}{\bibfnamefont{B.~D.} \bibnamefont{McCombe}},
  \bibinfo{journal}{Phys. Rev. B} \textbf{\bibinfo{volume}{29}},
  \bibinfo{pages}{3180} (\bibinfo{year}{1984}).







\end{thebibliography}


\end{document}